\documentclass[pre]{revtex4}
\usepackage[dvips]{graphicx}
\begin{document}
\title{Branching Patterns and Stepped Leaders in an Electric-Circuit Model for Creeping Discharge}
\author{Hidetsugu Sakaguchi and Sahim M. Kourkouss}
\affiliation{Department of Applied Science for Electronics and Materials,
Interdisciplinary Graduate School of Engineering Sciences, Kyushu
University, Kasuga, Fukuoka 816-8580, Japan}
\begin{abstract}
We  construct a two-dimensional electric circuit model for creeping discharge.
Two types of discharge, surface corona and surface leader, are modeled by a
 two-step function of conductance. Branched patterns of surface leaders surrounded by the surface corona appear in numerical simulation. The fractal dimension of branched discharge patterns is calculated by changing voltage and capacitance. We find that surface leaders often grow stepwise in time, as is observed in lightning leaders of thunder.
\end{abstract} 
\maketitle
\section{Introduction}
Various branching patterns such as dendritic crystals, river networks and blood vessel networks have been intensively studied from a view point of fractals.
\cite{rf:1} Discharge patterns such as lightning are also typical branching patterns.\cite{rf:2} In creeping discharge, a strong voltage is applied at a point electrode on the surface of dielectric materials. Discharge patterns appear around the point electrode at the interface between the dielectric materials and the surrounding gas or liquid.  In a suitable range of voltages, densely branched patterns are observed, that are called Lichtenberg figures. 
Fractal analysis was performed for the Lichtenberg figures obtained in experiments by several groups.\cite{rf:3,rf:4} Kebbabi and Bernounal found that the fractal dimension decreases with increasing thickness of dielectric materials.

The DLA model is a simple stochastic model that can generate a fractal branching pattern.\cite{rf:5} 
The fractal dimension is 1.71 in the DLA model. However, there are a variety of discharge patterns, which have various fractal dimensions $D_f$. 
In typical patterns in lightning discharge, the fractal dimension is rather small, e.g., $D_f\sim 1.2$.  For densely branched Lichtenberg figures, the fractal dimension is $D_f\sim 1.8$.  The DLA model was generalized for application to the various discharge patterns. Niemeyer et al. proposed the $\eta$ model in which  the growth probability is proportional to $E^{\eta}$, where $E$ is the electric field.\cite{rf:6}  Wiesmann and Zeller proposed another model in which the growth probability is proportional to $E-E_c$ for $E>E_c$, where $E_c$ is the threshold value of discharge.\cite{rf:7} Kuoershtokh et al. proposed another stochastic model in which there is a transition from  a weak discharge (streamer) to a strong discharge (leader).\cite{rf:8}  
We proposed a deterministic electric circuit model of branching patterns in the barrier discharge where an alternating voltage is applied between 
dielectric electrodes.\cite{rf:9}  

Various complicated dynamics are observed in the growth process of discharge patterns. For creeping discharge, growth dynamics was studied experimentally using a high-speed camera.  When the electric voltage $V$ is increased over its threshold value, surface corona appears locally around the central electrode.  The corona discharge is a weak discharge that appears locally around a sharp electrode.  When $V$ is beyond a second threshold, surface leaders appear, which make a branching pattern. A surface leader shows a stronger discharge than the surface corona.  The surface corona grows further around the tips of surface leaders. 
  The length of surface leaders increases with $V$.\cite{rf:10}  When a surface leader reaches the outer electrode, a flashover occurs, and a large current flows, leading to dielectric breakdown. 

The growth of  surface leaders is not always smooth in time, but intermittent in time.\cite{rf:11} 
A similar intermittent growth of the leaders is well known in lightning discharge.\cite{rf:2}  When a leader goes down from a thunder cloud to the ground, it moves in steps of about 30 m with a pause of about 40 ms between steps.  This leader is called the stepped leader.  When the leader reaches the ground, a flashover occurs and a strong flash called a return stroke appears, which we observe as lightning. The origin of stepped leaders is not completely understood.
There are several qualitative theories of such stepped leaders. 
Bruce pointed out the importance of the transition from a weak glow discharge to a strong arc discharge.\cite{rf:12} Kumar and Nagabhushana proposed a simulation model of stepped leaders based on a complicated electric breakdown process.\cite{rf:13}  

In this paper, we propose a simple deterministic electric circuit model composed of resistors and capacitors for the creeping discharge and perform numerical simulation. We have assumed a two-step function for the conductance, which exhibits a transition from a weak discharge to a strong discharge.    
We show that densely branched patterns appear when the capacitance is relatively large.  We find an intermittent growth of the discharge pattern, which is similar to the stepped leader.

\section{Electric Circuit Model}
Our electric circuit model is composed of capacitors and resistors whose conductance changes with the voltage. The conductance $\sigma$ of the resistance is assumed to change stepwise as shown in Fig.~1(a).  The conductance $\sigma$ increases rapidly owing to the ionization, when  voltage is increased beyond its threshold value and the discharge sets in. Taking another transition between the surface corona to the surface leader into account, we assume that there is another  threshold and the conductance $\sigma$ increases further at the second threshold. We study a dynamical system for $\sigma$ and the electric potential $V$. The time evolution of $\sigma$ is assumed to obey 
\begin{eqnarray}
\tau\frac{d\sigma}{dt}&=&\sigma_1-\sigma, \;\;\;{\rm for}\;\; |\Delta V|<V_{c1},\nonumber\\
\tau\frac{d\sigma}{dt}&=&\sigma_2-\sigma, \;\;\;{\rm for}\;\; V_{c1}<|\Delta V|<V_{c2},\nonumber\\
\tau\frac{d\sigma}{dt}&=&\sigma_3-\sigma, \;\;\;{\rm for}\;\; |\Delta V|>V_{c2},
\end{eqnarray}
where $\Delta V$ is the electric potential difference, $V_{c1}$ and $V_{c2}$ are the two threshold values, $\sigma_{1,2,3}$ are the stepped values of conductance, and $\tau$ is a time constant. We further assume that the threshold values depend on $\sigma$ as  $V_c=V_{c0}-\alpha\sigma$, which makes the transitions discontinuous transitions accompanying hysteresis.  
Hysteresis is often observed in the ionization transition or a transition between different types of discharge states. 

The resistors are set on edges of a triangular lattice as shown in Fig.~1(b). The conductance of the resistor between two sites $(i,j)$ and $(i^{\prime},j^{\prime})$ is denoted as $\sigma_{i,j,i^{\prime},j^{\prime}}$, and the electric potential at the $(i,j)$ site is denoted as $V_{i,j}$. The conductance $\sigma_{i,j,i^{\prime},j^{\prime}}$ obeys eq.~(1) for $\Delta V=V_{i,j}-V_{i^{\prime},j^{\prime}}$.  That is, 
\begin{eqnarray}
\tau\frac{d\sigma_{i,j,i^{\prime},j^{\prime}}}{dt}&=&\sigma_1-\sigma, \;\;\;{\rm for}\;\; |\Delta V_{i,j,i^{\prime},j^{\prime}}|<V_{c10}-\alpha\sigma_{i,j,i^{\prime},j^{\prime}},\nonumber\\
\tau\frac{d\sigma_{i,j,i^{\prime},j^{\prime}}}{dt}&=&\sigma_2-\sigma, \;\;\;{\rm for}\;\; V_{c10}-\alpha\sigma_{i,j,i^{\prime},j^{\prime}}<|\Delta V_{i,j,i^{\prime},j^{\prime}}|<V_{c20}-\alpha\sigma_{i,j,i^{\prime},j^{\prime}},\nonumber\\
\tau\frac{d\sigma_{i,j,i^{\prime},j^{\prime}}}{dt}&=&\sigma_3-\sigma, \;\;\;{\rm for}\;\; |\Delta V_{i,j,i^{\prime},j^{\prime}}|>V_{c20}-\alpha\sigma_{i,j,i^{\prime},j^{\prime}}.
\end{eqnarray}
Each site is connected to the ground with a capacitor of capacitance $C$ as \begin{figure}[tbp]
\begin{center}
\includegraphics[height=3.5cm]{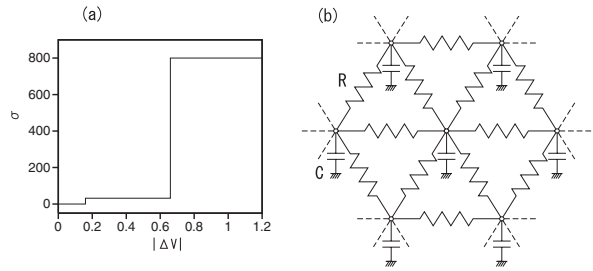}
\end{center}
\caption{(a) Schematic figure of the conductance $\sigma$ as a function of electric potential difference $\Delta V$. (b) Schematic figure of an electric circuit on a triangular lattice.}
\label{f1}
\end{figure}
shown in Fig.~1(b). The capacitors represent the dielectric material on which the creeping discharge occurs. The electric potential $V$ of the central (1+6) sites is set to be $V_0$ and the electric potential outside of a hexagonal area of side length $L$ is set to be 0. The electric potential $V_{i,j}$ at $(i,j)$ site obeys
\begin{equation}
C\frac{dV_{i,j}}{dt}=-\sum_{i^{\prime},j^{\prime}}\sigma_{i,j,i^{\prime},j^{\prime}}(V_{i,j}-V_{i^{\prime},j^{\prime}}),
\end{equation}
from the conservation law of the electric charge. 
The summation is taken over the nearest neighbor sites $(i^{\prime},j^{\prime})$ of $(i,j)$.
We show several numerical results for $L=75, V_{c10}=0.16, V_{c20}=0.66,\sigma_1=0.032,\sigma_2=32, \sigma_3=800, \tau=0.1$, and $\alpha=0.02$. $C$ and $V_0$ are changed as control parameters. However, similar results were also obtained for different parameter values. 

\begin{figure}[tbp]
\begin{center}
\includegraphics[height=3.5cm]{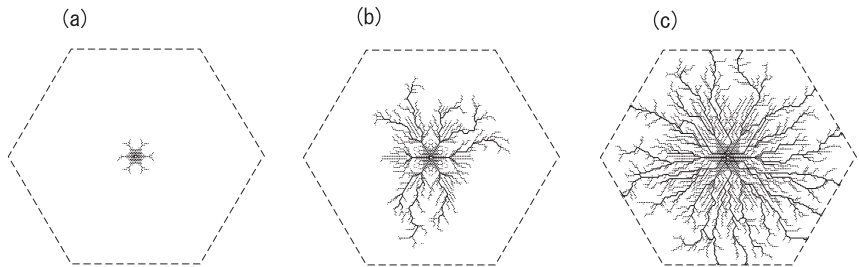}
\end{center}
\caption{Discharge patterns at (a) $V_0=0.45$, (b) 0.6, and (c) 0.7 for $C=0.035$. Discharged edges whose electric potential difference is beyond the first threshold are plotted using dots. The solid lines denote discharged edges whose electric potential difference is beyond the second threshold. The inner and outer electrodes are drawn with dashed lines.
}
\label{f2}
\end{figure}
\begin{figure}[tbp]
\begin{center}
\includegraphics[height=4.cm]{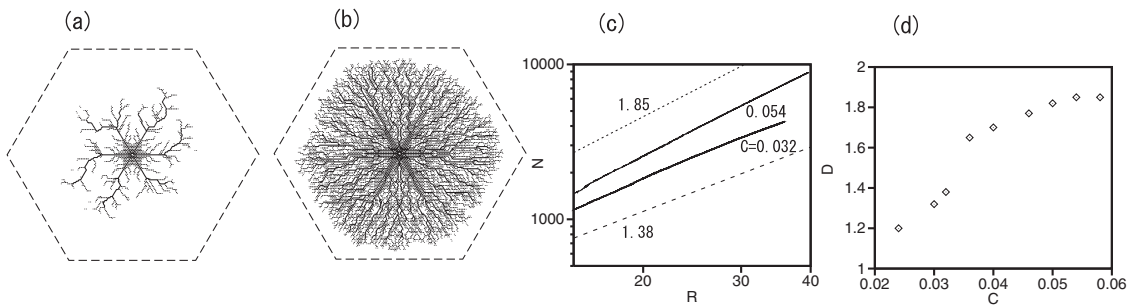}
\end{center}
\caption{Discharge patterns at (a) $C=0.024$ and (b) 0.054 for $V_0=0.7$. 
(c) Double logarithmic plot of $R(t)$ vs. $N(t)$ for $C=0.032$ and $0.054$.
Two straight lines denote $N\sim R^{1.85}$ and $N\sim R^{1.38}$. (d) Fractal dimension $D$ of discharged patterns for several $C$'s. 
}
\label{f3}
\end{figure}

\section{Branching patterns}
Figure 2 shows discharge patterns at three different values of $V_0$ for $C=0.035$. 
The initial conditions are $\sigma_{i,j,i^{\prime},j^{\prime}}=0.001+0.0001\times r$, where $r$ is a random number between 0 and 1, and $V_{i,j}=0$. 
The initial conditions are slightly random, but there is no randomness in system parameters.  
The corona discharge appears for $V_0\ge 0.23$ in our system. For $V_0\le 0.22$, all $\sigma_{i,j,i^{\prime},j^{\prime}}$'s approach $\sigma_1$, because the electric potential difference is below the first threshold at all sites. 
Figure 2(a) shows a discharge pattern at $V_0=0.45$. The inner and outer hexagonal electrodes are drawn using dashed lines. (The inner electrode is hardly seen in these plots because it is too small.)  The discharge pattern stops growing at a finite time.  In this discharge pattern, $\sigma_{i,j,i^{\prime},j^{\prime}}$ takes a value of $\sigma_2$ in the corona region. Corona discharge is a discharge state that appears only around a sharp electrode.  For $V_0\ge 0.50$, the electric potential difference goes beyond the second threshold, a stronger discharge state appears, that is,  a few surface leaders grow from the inner electrode. 
The surface corona appears around the surface leaders.  
Figure 2(b) shows a discharge pattern at $V_0=0.6$. A random discharge pattern appears even in the uniform system. 
This is probably because chaotic dynamics or dynamical instability is involved in our deterministic system, similarly to the interface instability in crystal growth~\cite{rf:14}.  The discharge pattern reaches the outer electrode with $V=0$ for $V_0\ge 0.61$ in this configuration. Figure 2(c) is a discharge pattern at $V_0=0.7$, which represents a flashover. 

The degree of ramification of branched patterns increases, as the capacitance $C$ is increased.
Figures 3(a) and 3(b) show discharge patterns at (a) $C=0.024$ and (b) 0.054 for $V_0=0.7$. For the quantitative analysis of the branched patterns, we have calculated the radius of gyration as
\begin{equation}
R(t)=\sqrt{\sum_i (x_i^2+y_i^2)/N(t)},
\end{equation}
where the summation is taken over the discharged sites whose electric potential difference is beyond the first threshold,  $x_i$ and $y_i$ are the $x$ and $y$ coordinates of the discharged site,  and $N$ is the total number of discharged sites at time $t$. Figure 3(c) shows double logarithmic plots of the relation of $R(t)$ and $N(t)$ for $C=0.032$ and 0.054. The fractal dimension is evaluated from the slope of the relation of $\ln N(t)$ vs $\ln R(t)$ as $D\sim 1.38$ for $C=0.024$, and as $D\sim 1.85$ for $C=0.054$. Figure 3(d) shows the fractal dimension $D$ as a function of $C$. As $C$ is increased, the discharge pattern is more densely branched and the fractal dimension increases, although the discharge pattern stops finally and the fractal dimension cannot be precisely evaluated using a sufficiently large pattern.  In the experiment by Kebbabi and Berounal, the fractal dimension becomes smaller as the thickness $d$ of the dielectric material is increased. The capacitance of the dielectric material is inversely proportional to the thickness, i.e., $C\propto 1/d$. Therefore, our numerical result is qualitatively consistent with their experimental result.  

\begin{figure}[tbp]
\begin{center}
\includegraphics[height=4.cm]{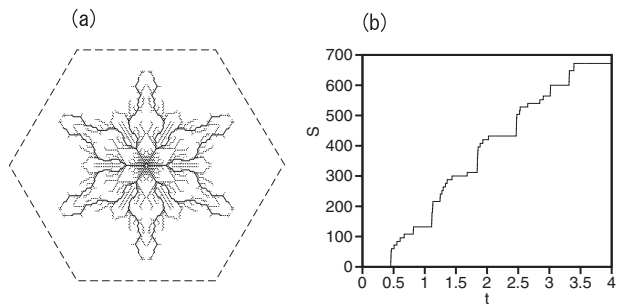}
\end{center}
\caption{(a) Discharge pattern under a uniform initial condition at $C=0.035$ and $V_0=0.6$. (b) Time evolution of the total number $S(t)$ of edges included in the surface leaders. 
}
\label{f4}
\end{figure}
\begin{figure}[tbp]
\begin{center}
\includegraphics[height=4.cm]{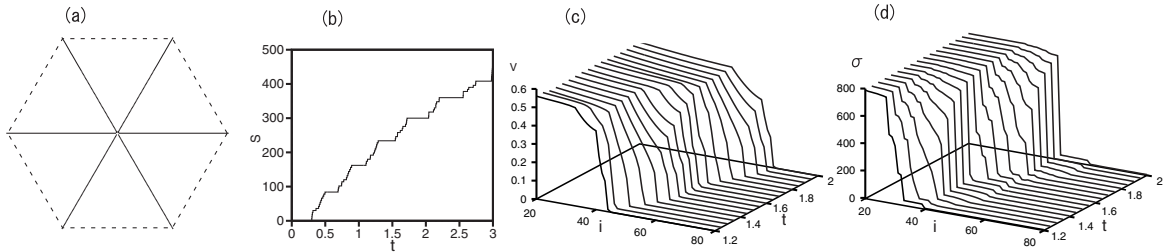}
\end{center}
\caption{(a) Discharge pattern from a uniform initial condition at $C=0.035$ and $V_0=0.6$. The discharge occurs only on  six straight lines. The second threshold is $V_{c20}=0.64$. 
(b) Time evolution of the total number $S(t)$ of edges included in the surface leaders.  (c) Time evolution of the electric potential $V(i,j)$ on the central line $j=L/2$.  (d) Time evolution of the conductance $\sigma_{i,j,i+1,j}$ on the central line $j=L/2$.
}
\label{f5}
\end{figure}
\section{Stepped Leaders}
In the time evolution of the discharge pattern, the surface leaders and corona discharge interact with each other. That is, the surface leaders appear on some selected edges in the corona region,  with a certain time delay after the appearance of corona discharge.  The corona discharge further grows around the tips of the surface leaders. In the numerical simulations in the previous section, we have used a slightly random initial conditions for $\sigma_{i,j,i^{\prime},j^{\prime}}$'s. 
If the initial conditions are uniform, a symmetrical pattern appears. Figure 4(a) shows such a hexagonally symmetric discharge pattern at $C=0.035$ and $V_0=0.6$. The initial condition is $\sigma_{i,j,i^{\prime},j^{\prime}}=0.01$ for all edges. Figure 4(b) shows the time evolution of the number $S$ of edges where the electric potential difference is beyond the second threshold. The number $S$ is proportional to the total length of the surface leaders. Figure 4(b) shows that the surface leaders grow stepwise in time. 
That is, the surface leaders stop and go intermittently. 
For example, the surface leaders grow rapidly at $t\sim 0.45$ from the inner electrode and stop at $t\sim 0.82$. 
This behavior is similar to that of the stepped leaders in lightning discharge. 
Even under the previous random initial conditions, this type of stepped leader is observed for each branch, but the total number $S(t)$  exhibits no such  clear stepped  behavior except for the first step from $S=0$, because the timing of the stop-and-go is different for each branch.

To understand the behavior of the stepped leaders, we have simulated an even simpler model system where only the edges on the six straight lines  can be discharged, as shown in Fig.~5(a).  Figure 5(b) shows the time evolution of $S(t)$ in this system for $C=0.035$, $V_0=0.6$, $V_{c10}=0.16$, and $V_{c20}=0.64$. The intermittent behavior is observed for $0.58\le V_{c20}\le 0.65$, when other parameters are fixed. At $V_{c20}=0.66$, the surface leaders do not grow in this system. For $V_{c20}\le 0.57$, the surface leaders move almost steadily.   Figure 5 shows that the branching is not directly related to the origin of the stepped leaders, because no branching occurs in this simple system. Figure 5(c) shows a quasi-3D plot of the time evolution of $V(i,j)$ on the central line $j=L/2$ for $20\le i\le 80$ between $1.2\le t\le 2$, and Fig.~5(d) show a quasi-3D plot of the time evolution of $\sigma_{i,j,i+1,j}$ on the central line $j=L/2$. The behavior of the stepped leaders is clearly observed in Fig.~5(d). The interaction between the corona discharge, surface leader, and electric potential is important for the intermittent motion. When $V_{c20}=V_{c10}$, the difference between the two types of discharge disappears, and then we have observed no intermittent motion.  These results are qualitatively consistent with previous observations by Shimazaki et al.\cite{rf:11} and Bruce\cite{rf:12}.

\section{Summary and Discussion}
We have constructed a simple deterministic electric circuit model for creeping discharge. We have found that random branching patterns appear, even though the system parameters are uniform. We have found that more densely branched patterns appear as the capacitance $C$ increases, which is qualitatively consistent with the experimental result. We have numerically found the behavior of stepped leaders, which seems to originate from the interaction with the corona discharge and  surface leaders, although the detailed theoretical analysis of the stepped leaders is left for future study. 

A similar intermittent growth is observed in other systems. 
The stick-slip motion in the frictional oscillation is a typical example. 
The threshold by the maximum static friction plays an important role in the stick-slip motion.  Wakita et al. found a periodic growth in a bacterial colony.\cite{rf:15}  Some threshold values of bacterial density seem to play an important role in  periodic growth. 
Nakanishi et al. found a periodic growth in the electrodeposition of Sn,\cite{rf:16} and Sakaguchi et al. proposed a coupled map lattice model of the periodic growth.\cite{rf:17} The transition from  a facet growth  to a diffusion-limited growth is induced by a certain threshold voltage in this electrodeposition system. We think that there is some analogy in these systems, in that there is a transition between different states via some threshold values.


\begin{thebibliography}{99}
\bibitem{rf:1} B.~B.~Mandelbrot: {\it The Fractal Geometry of Nature} (Freeman, San Francisco, 1982).
\bibitem{rf:2} M.~A.~Uman: {\it The Lightning Discharge} (Academic, San Diego, 1987).
\bibitem{rf:3} T.~Ficker: J. Phys. D {\bf 32} (1999) 219. 
\bibitem{rf:4} L.~Kebbabi and A.~Berounal: J. Phys. D {\bf 39} (2006) 177. 
\bibitem{rf:5} T.~A.~Witten and L.~M.~Sander: Phys. Rev. Lett. {\bf 47}  (1981) 1400
\bibitem{rf:6} L.~Niemeyer, L.~Pietronero, and H.~J.~Wiesmann: Phys. Rev. Lett. {\bf 52} (1984) 1033.
\bibitem{rf:7} H.~J.~Wiesmann and H.~R.~A.~Zeller: J. Appl. Phys. {\bf 60} (1986) 1770.
\bibitem{rf:8} A.~L.~Kupershtokh, V.~Charalambakos, D.~Agoris, and D.~I.~Karpov:
 J. Phys. D {\bf 34} (2001) 936.
\bibitem{rf:9} H.~Sakaguchi: Prog. Theor. Phys. {\bf 117} (2007) 219.
\bibitem{rf:10} M.~Toepler: Ann. Phys. {\bf 21} (1906) 193.
\bibitem{rf:11} T.~Shimazaki, I.~Tsuneyasu, and M.~Akasaki: IEEJ Trans. on Electric. Electron. Eng. {\bf 99} (1979) 527 (in Japanese).
\bibitem{rf:12} C.~E.~R.~Bruce: Proc. R. Soc. London {\bf A183} (1944) 228. 
\bibitem{rf:13} U.~Kumar and G.~R.~Nagabhushana: IEEE Proc. Sci. Meas. Technol.  {\bf 147} (2000) 56.  
\bibitem{rf:14} H.~Sakaguchi and S.~Tokunaga: Physica D {\bf 205} (2005) 222. 
\bibitem{rf:15} J.~Wakita, H.~Shimada, H.~Itoh, T.~Matsuyama, and M.~Matsushita: J. Phys. Soc. Jpn. {\bf 70} (2001) 911.
\bibitem{rf:16} S.~Nakanishi, K.~Fukami, T.~Tada, and Y.~Nakato: J. Am. Chem. Soc. {\bf 126} (2004) 9556.
\bibitem{rf:17} H.~Sakaguchi, T.~Yoshida, S.~Nakanishi, K.~Fukami, and Y.~Nakato: J.Phys. Soc. Jpn. {\bf 75} (2006) 114002. 
\end{thebibliography}
\end{document}